\documentclass{aa}  
\usepackage{txfonts}
\usepackage{hyperref} 

\usepackage{array} 
\usepackage{tabularx}

\newcommand{\dwnine}{DE0823$-$49}
\begin{document} 
   \title{\dwnine\ is a juvenile binary brown dwarf at 20.7 pc\thanks{Based on observations made with ESO telescopes at the La Silla Paranal Observatory under programme IDs 086.C-0680, 088.C-0679, 090.C-0786, and 092.C-0202.}}
   
\author{J. Sahlmann\inst{1}\fnmsep\thanks{ESA Research Fellow}
\and A. J. Burgasser	\inst{2}\fnmsep\thanks{Visiting professor at the Instituto de Astrof\'isica de Canarias (IAC), La Laguna, Tenerife, Spain}\fnmsep\thanks{Visiting Astronomer at the Infrared Telescope Facility, which is operated by the University of Hawaii under Cooperative Agreement no. NNX-08AE38A with the National Aeronautics and Space Administration, Science Mission Directorate, Planetary Astronomy Program.}
		\and E. L. Mart\'in\inst{3} 
		\and P. F. Lazorenko\inst{4}
		\and D. C. Bardalez Gagliuffi\inst{2}
	        \and M. Mayor\inst{5} 
		\and D.~S\'egransan\inst{5}
		\and D. Queloz\inst{5,6} 
		\and S. Udry\inst{5}
		}	

\institute{European Space Agency, European Space Astronomy Centre, P.O. Box 78, Villanueva de la Ca\~nada, 28691 Madrid, Spain\\
		\email{Johannes.Sahlmann@esa.int}
		\and
		Center for Astrophysics and Space Science, University of California San Diego, La Jolla, CA, 92093, USA
		\and
		INTA-CSIC Centro de Astrobiolog\'ia, 28850 Torrej\'on de Ardoz, Madrid, Spain
		\and
		Main Astronomical Observatory, National Academy of Sciences of the Ukraine, Zabolotnogo 27, 03680 Kyiv, Ukraine
		\and
		Observatoire de Gen\`eve, Universit\'e de Gen\`eve, 51 Chemin Des Maillettes, 1290 Versoix, Switzerland
		\and
		University of Cambridge, Cavendish Laboratory, J J Thomson Avenue, Cambridge, CB3 0HE, UK
}

\date{Received 18 December 2014 / Accepted 28 May 2015} 

\abstract
{Astrometric monitoring of the nearby early-L dwarf \dwnine\ has revealed a low-mass companion in a 248-day orbit that was announced in an earlier work. Here, we present new astrometric and spectroscopic observations that allow us to characterise the system in detail. The optical spectrum shows \ion{Li}{i}-absorption indicative of a young age and/or substellar mass for the primary component. The near-infrared spectrum is best reproduced by a binary system of brown dwarfs with spectral types of {L1.5 $+$ L5.5} and effective temperatures of {$2150\pm100$ K and $1670\pm140$ K}. To conform with the photocentric orbit size measured with astrometry and the current understanding of substellar evolution, the system must have an age in the {80--500} Myr range. Evolutionary models predict component masses in the {ranges of $M_1\simeq0.028-0.063\,M_\sun$ and $M_2\simeq0.018-0.045\,M_\sun$ with a mass ratio of $q\simeq0.64-0.74$}. Multi-epoch radial velocity measurements unambiguously establish the three-dimensional orbit of the system and allow us to investigate its kinematic properties. \dwnine\ emerges as a rare example of a nearby brown dwarf binary with orbit, component properties, and age that are characterised well. It is a juvenile resident of the solar neighbourhood, but does not appear to belong to a known young association or moving group. } 

\maketitle

\section{Introduction}
Binary stars are vital for advancing our understanding of stellar formation and evolution. They provide us with the opportunity of direct mass measurements through orbital motion, and they are calibration systems for evolutionary models, which have to match the observations of a two-body system that presumably is co-eval and has near-identical composition. Very low-mass binary systems composed of brown dwarfs and/or ultracool dwarfs (spectral type M7 and later) are no exception and are in many ways more enlightening, given that brown dwarfs dim and cool substantially over time, yet they are less common than their higher-mass counterparts. The discovery and characterisation of ultracool dwarf binaries is thus a rare opportunity to yield observational input for the refinement of models, eventually leading to a better understanding of ultracool dwarf physics. 

The L dwarf \object{DENIS J082303.1-491201}, hereafter \dwnine, was targeted as part of an astrometric planet search \citep{Sahlmann:2014aa} and consequently identified as an astrometric binary \citep{Sahlmann:2013ab}. On the basis of features in its optical spectrum \citep{Phan-Bao:2008fr}, \cite{Sahlmann:2013ab} suggested that \dwnine\ is younger than the average field population, which corresponds to a primary at the hydrogen-burning mass limit orbited by a low-mass brown dwarf secondary. Here, we present follow-up spectroscopic observations that allow us to better constrain the system's properties, in particular the components' effective temperatures, the system age, and individual masses.

\section{Observations}
\subsection{VLT/FORS2 imaging}
We obtained two new astrometric epochs with the FORS2 instrument \citep{Appenzeller:1998lr} at ESO's Very Large Telescope (VLT), in addition to the data used for the orbit discovery. Details on the observation strategy can be found in \cite{Sahlmann:2013ab, Sahlmann:2014aa}. The photocentre computations for \dwnine\ are complicated by a close background star, which resulted in noticeable systematic errors for this object compared to other targets of the astrometric survey. Therefore a special model that extends the work of \cite{Lazorenko:2014aa} was devised for the reduction of \dwnine\ images. In this model, the seeing-dependent light contribution of the background star is taken into account.  A second improvement concerns the enhanced suppression of parameter correlations when reducing data of objects that show significant orbital motion. All data were re-reduced with the improved methods and used for the orbit adjustment.

\subsection{VLT/UVES spectroscopy}
We observed \dwnine\ on 2013 October 7 (MJD\footnote{Modified Julian date (MJD) is barycentric Julian date -- 2400000.5.} 56572.3332) with the red arm of UVES at the VLT \citep{Dekker:2000aa} using a $1\farcs2$ slit width, which provided a resolving power of $R$$\sim$33\,000, and the Dichroic 2 standard setup centred at 760 nm to cover the wavelength range of 565 - 931 nm. The exposure time was 2830 s and the observation took place with 1\farcs01 optical seeing at an airmass of 1.52. The spectrum was recorded on two separate chips (REDL and REDU) and was reduced using the ESO pipeline in standard setup. 

\subsection{IRTF/SpeX spectroscopy}
We observed \dwnine\ on 2013 November 24 (UT) with the SpeX spectrograph on the NASA Infrared Telescope Facility. (IRTF; \citealt{Rayner:2003aa}).  Conditions were clear but windy with poor seeing (1$\farcs$4 at $H$-band).  We used the SpeX prism mode with the 0$\farcs$5 slit aligned with the parallactic angle, yielding 0.8--2.45 ~$\mu$m spectra with an average resolution $\lambda/{\Delta}{\lambda}$ $\approx$ 120. Eight exposures of 120~s each were obtained at an airmass of 2.81, followed by observations of the A0\,V star \object{HD 83719} ($V$ = 7.64) at an airmass of 2.79. HeNeAr arc lamps and quartz lamp exposures were also obtained for dispersion and pixel response calibration. Data were reduced using SpeXtool version 3.4 \citep{Cushing:2004aa,Vacca:2003aa} following standard procedures for point-source extraction.

\subsection{Keck/NIRSPEC spectroscopy}
We obtained two high-resolution infrared spectra of \dwnine\ with the NIRSPEC echelle spectrograph on the Keck II telescope \citep{2000SPIE.4008.1048M} on 2014 April 14 and December 8. Conditions on both nights were clear with 0$\farcs$6 and 0$\farcs$9 seeing at $K$-band, respectively. We used the N7 order-sorting filter and 0$\farcs$432-wide slit to obtain 2.00--2.39~$\mu$m spectra over orders 32--38 with $\lambda/\Delta\lambda$ = 20\,000 ($\Delta{v}$ = 15~{km~s$^{-1}$}) and dispersion of 0.315~{\AA}~pixel$^{-1}$. Two dithered exposures of 600~s each were obtained at an airmass of 2.78, followed with observations of the A0\,V star \object{HD 87363} ($V$ = 6.11). Flat field and dark frames were obtained at the start of the night with the same instrument setting. 

\section{Analysis and interpretation}

\begin{table}
\caption{Updated astrometric parameters of the \dwnine\ system.}    
\label{table:1}      
\centering                       
\begin{tabular}{l c r} 
\hline\hline    
$\Delta\alpha^{\star}_0$& (mas)&   $-235.04_{   -0.16}^{+    0.16}$ \\ [2pt]
$\Delta\delta_0$& (mas)&    $-13.19_{   -0.30}^{+    0.34}$ \\[2pt]
$\varpi$& (mas)&     48.27$_{   -0.12}^{+    0.12}$\\[2pt]
$\mu_{\alpha^{*}}$& (mas yr$^{-1}$) &   $-154.92_{   -0.06}^{+    0.06}$ \\[2pt]
$\mu_{\delta}$& (mas yr$^{-1}$)&      7.99$_{   -0.06}^{+    0.06}$ \\[2pt]
$e$ & &      0.36$_{   -0.04}^{+    0.04}$\\[2pt]
$\omega$ & (deg)&     41.8$_{   -4.9}^{+    4.4}$\\[2pt]
$P$& (day)&    247.75$_{   -0.64}^{+    0.64}$\\[2pt]
$\lambda_{\rm Ref}$ & (deg)&   -365.8$_{   -2.1}^{+    2.0}$\\[2pt]
$\Omega$ & (deg)&    $-13.8\pm2.0$\\ [2pt] 
$i$ & (deg)&     52.2$_{   -1.5}^{+    1.4}$\\[2pt]
$\alpha$& (mas)&      4.62$_{   -0.11}^{+    0.12}$ \\[2pt]
$\rho$ & (mas)&     20.8$_{   -1.8}^{+    1.7}$\\[2pt]
$d$& (mas)&    $-26.2\pm1.5$\\ 
$s_\alpha$ & (mas) & {$0.15_{-0.10}^{+0.09}$}\\[2pt] 
$s_\delta$ & (mas) & {$0.12_{-0.08}^{+0.09}$}\\[3pt] 
\multicolumn{3}{c}{Derived and additional parameters}\\[1pt]
$T_\mathrm{Ref}$ &(MJD)& {55926.823928}\\[2pt]
$\Delta\varpi$& (mas)&     $-0.06 \pm0.04$ \\[2pt] 
$\varpi_\mathrm{abs}$ & (mas)&     $48.33\pm 0.14$ \\[2pt] 
Distance & (pc)&     $20.69 \pm0.06$\\[2pt] 
\multicolumn{2}{l}{Number of epochs / frames}  & {16 / 334}\\
$\sigma_{O - C, \mathrm{Epoch}}$ & (mas)  & {0.176}\\ 
\hline
\end{tabular}
\tablefoot{Parameter values are the median of the marginal parameter distributions and uncertainties represent 1$\sigma$-equivalent ranges. The proper motions are not absolute and were measured relative to the local reference frame. The inclination is measured relative to the sky plane.}
\end{table}

Because of the new astrometric measurements and the improved reduction method, we re-analysed the astrometric data that comprise 16 epochs over a timespan of 831 days. The analysis methods are described in detail in \cite{Sahlmann:2013ab} and consist of a genetic algorithm followed by a Markov-Chain Monte Carlo (MCMC) code. The updated fit parameters and their confidence intervals are reported in Table \ref{table:1}, where $\Delta\alpha^{\star}_0$ and $\Delta\delta_0$ are relative offsets to the target's position at the reference date $T_\mathrm{Ref}$ taken as the arithmetic mean of the observation dates, $\varpi$ is the relative parallax, $\mu_{\alpha^\star}$ and $\mu_\delta$ are the proper motions, $e$ is the eccentricity, $\omega$ is the argument of periastron, $P$ is the orbital period, $\lambda_{\rm Ref}$ is the mean longitude at $T_\mathrm{Ref}$, $\Omega$ is the ascending node, $i$ is the orbit's inclination, and $\alpha$ is the semi-major axis of the photocentric orbit. The parameters $\rho$ and $d$ model the differential chromatic refraction and $s_\alpha$ and $s_\delta$ are nuisance parameters. The parallax correction $\Delta\varpi$ was determined in \cite{Sahlmann:2013ab} and yields the absolute parallax $\varpi_\mathrm{abs}$. The time of periastron passage $T_0$ can be retrieved via the mean anomaly 
\begin{equation}\label{eq:meanano}
M= \lambda - \omega = 2\pi \left( \frac{t}{P} -\phi_0  \right)\;\; \Rightarrow \;\;T_0 = T_\mathrm{Ref} - P \frac{M_\mathrm{Ref}}{2\pi},
\end{equation}
where t is time and $\phi_0 = T_0/P$ is the phase at periastron.
  
The most significant change is the reduction of the residual scatter from 0.330 milli-arcseconds (mas) \citep{Sahlmann:2013ab} to $\sigma_{O - C, \mathrm{Epoch}}=0.176$ mas, which reflects the improved data reduction. Consequently, the confidence intervals of most parameters are smaller as well. A notable change in the orbital parameters is a smaller inclination ($56.6\pm2.0$\degr\ before, $52.2\pm1.5$\degr\ in this work). Other parameters, in particular the parallax,  period, eccentricity, and photocentric semi-major axis $\alpha$, remain unchanged when accounting for the uncertainties.

\begin{figure}[h!]
\centering
\includegraphics[width = \linewidth]{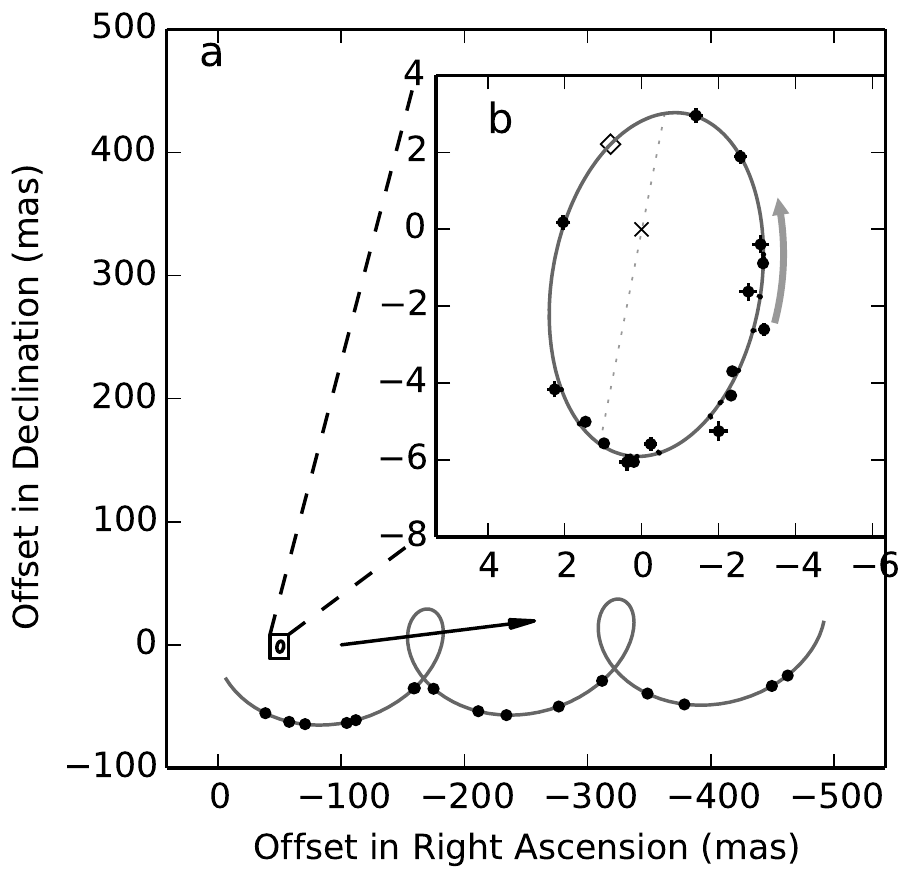}
\caption{Astrometric motion of \dwnine~and its photocentric orbit, updated from \cite{Sahlmann:2013ab}. Panel \textbf{a} shows proper and parallactic motion relative to the field of reference stars. Panel \textbf{b} is a close-up of the photocentric orbit caused by the gravitational pull of the orbiting brown dwarf. Observations with uncertainties and the best-fit model are shown as black circles and grey curve, respectively.}
\label{fig:orbit} 
\end{figure}

\subsection{Radial velocity measurements}
\begin{figure}[h!]
\centering
\includegraphics[width = \linewidth, trim=10mm 35mm 10mm 20mm, clip]{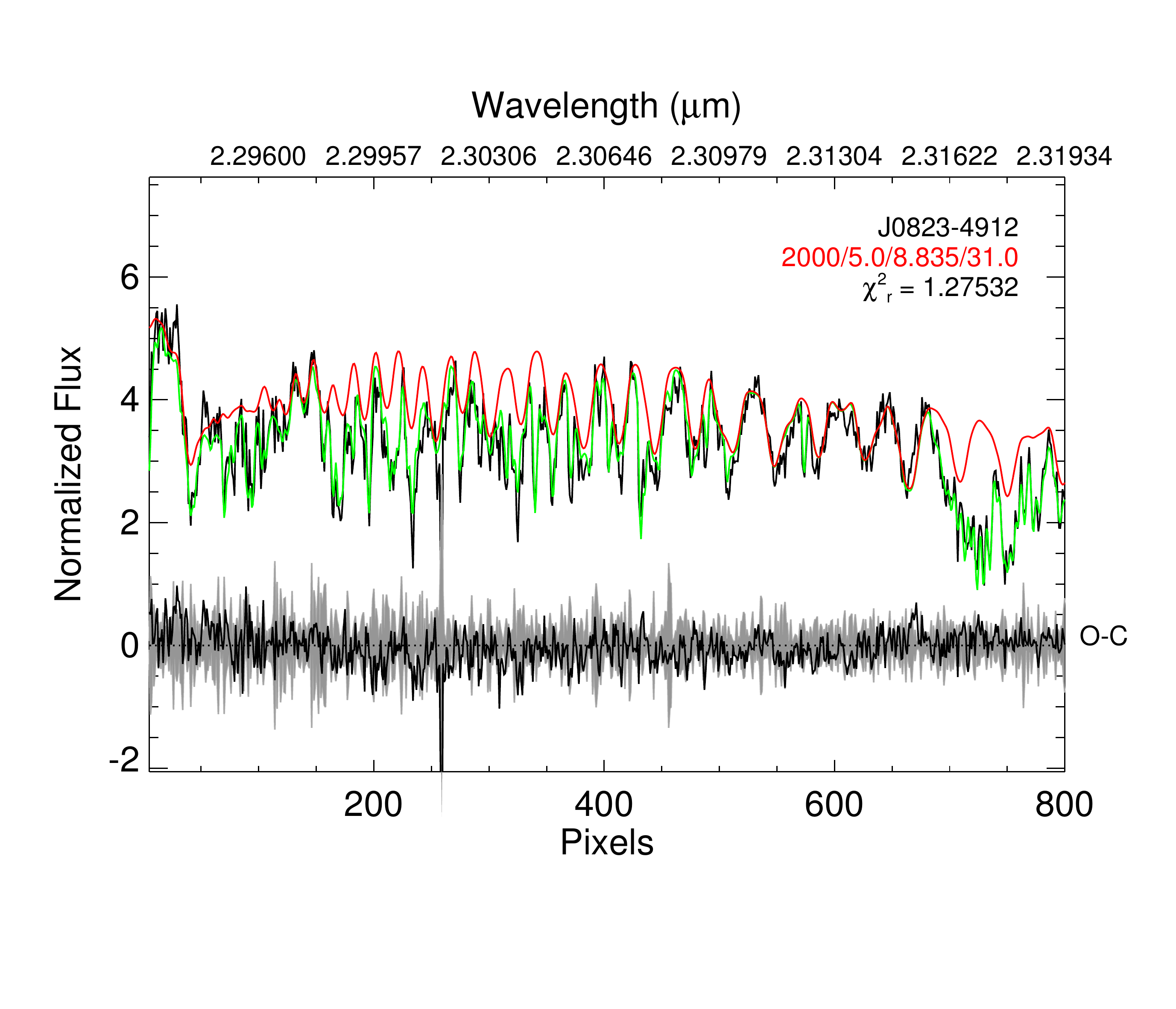}
\caption{High-resolution K-band spectrum of \dwnine\ obtained with NIRSPEC (black line) compared to a best-fit spectral (red line) and spectral plus telluric model (green line).  Difference between data and model is plotted in black at bottom; the uncertainty spectrum is indicated in grey. The values listed below the target name on the top right are $T_\mathrm{eff}$/$\log g$/RV/$v \sin{i}$.} 
\label{fig:keck1}
\end{figure}

The Keck/NIRSPEC spectra were optimally extracted using a modified version of the REDSPEC package; the spectra on each night had a S/N$\sim$12. These data were then forward-modelled using a custom MCMC implementation of the method described in \cite{Blake:2010lr}. We used the Solar atlas of \cite{Livingston:1991aa} to model telluric absorption and the BT-Settl atmosphere models \citep{Allard:2011aa} to model the spectrum of \dwnine.  A {$T_\mathrm{eff}$} = 2000~K, {$\log{g}$} = 5.0 (cgs) model provided the best fit. Figure \ref{fig:keck1} shows that the extracted spectrum is a good fit to the final model. The distribution of chain values yields the mean heliocentric radial velocities and projected rotational velocities listed in Table \ref{table:3}.

We also measured the radial velocity on the UVES spectrum using four strong atomic lines clearly visible in the data, namely \ion{Rb}{i} at 780.0268 nm, \ion{Rb}{i} at 794.7603 nm, \ion{Cs}{i} at 852.1132 nm, and \ion{Cs}{i} at 894.3474 nm. The rest wavelengths for these lines were obtained from the NIST Atomic Spectra Database (version 5.1; \citealt{Kramida:2012aa}). The wavelengths of the centroid of the lines in the observed spectrum were derived using line profile fitting with the task \texttt{splot} in IRAF.  Heliocentric correction was applied using the IRAF task \texttt{rvcorrect} and the information of Julian Date provided in the FITS header. This procedure was checked using archival UVES spectra for the L1 dwarf \object{2MASS J10484281+0111580} from ESO program ID 078.C-0025(A) (PI A. Reiners). With our method we obtained a heliocentric radial velocity of 21.9 $\pm$ 1.4 km/s, which is consistent with the published value of 24.0 $\pm$ 1.1 km/s \citep{Seifahrt:2010fj}. No significant (i.e. larger than our uncertainties) radial velocity offset is expected to exist between the NIRSPEC and UVES measurements, as these two instruments have been shown to give comparable results for the brown dwarf \object{LP944-20}, although with larger scatter in the UVES based measurements \citep{Martin:2006aa}. For \dwnine, we determined a heliocentric radial velocity of $+6.5 \pm 1.5$ km/s. 

A fourth radial velocity of $+12.5 \pm 2.4$ km/s was obtained from observations with the Magellan Echellette Spectrograph (MagE) in 2009. This measurement will be discussed in Burgasser et al. (in prep.). Using 32 sources that were both observed with MagE and with UVES \citep{Seifahrt:2010fj,Reiners:2009kx}, we determined an average velocity offset between both instruments and reduction methods of $-0.29\pm0.59$ km/s. It is small and not significant, especially when compared to the uncertainty of the MagE measurement of \dwnine, and we therefore neglected it in our analysis.

\begin{figure}[h!]
\centering
\includegraphics[width = \linewidth, trim=15mm 0 10mm 0, clip=true]{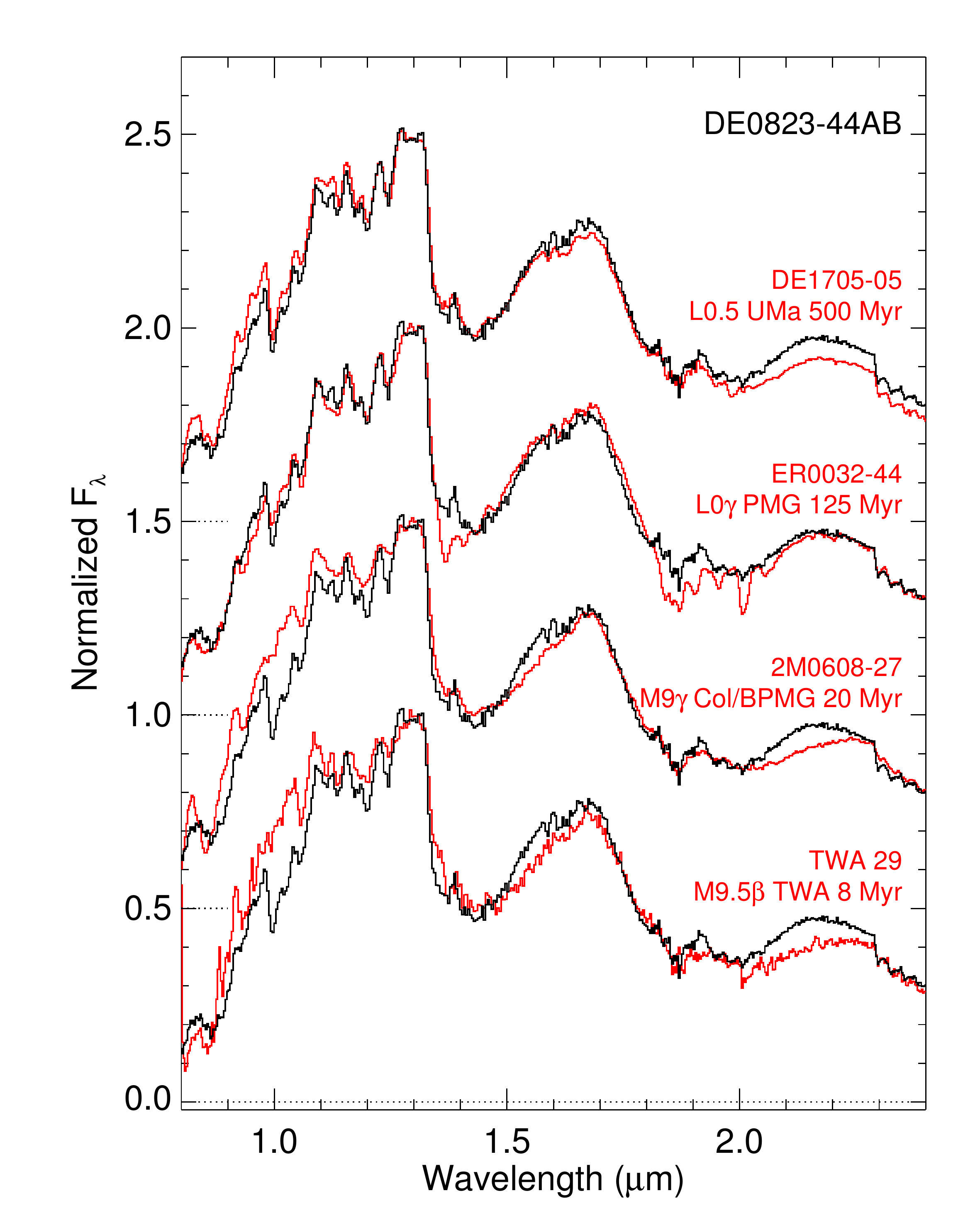}
\caption{SpeX spectrum of \dwnine\ (black) compared to four sources (red curves) with similar spectral types and well-constrained ages. Source identifier, optical spectral type, and the approximate age are indicated next to every spectrum. Cluster memberships (UMa = Ursa Majoris, PMG = Pleiades Moving Group, BPMG = Beta Pic Moving Group, TWA = TW Hydrae Association) are from \cite{Allers:2013aa}.}
\label{fig:spex1}
\end{figure}

\subsection{Spectroscopic age indicators} 
Figure \ref{fig:spex1} shows the reduced SpeX spectrum of \dwnine\ compared to four sources with reasonable age estimates and similar spectral types that were chosen to best match the $1.3-1.4\,\mu$m water band: \object{TWA 29} ($\sim$8 Myr, i.e.\ the age of the TW Hydrae Association), \object{2MASS J06085283-2753583} (2M0608$-$27, $\sim$20 Myr, i.e.\ the age of the Beta Pic Moving Group), \object{EROS-MP J0032-4405} (EROS 0032$-$44, $\sim$100 Myr, i.e.\ the age of the Pleiades Moving Group or local association) and \object{DENIS J170548.3-051645} (DE1705-05, $\sim$500 Myr, i.e.\ the age of the Ursa Majoris Moving Group).  The comparison sources are from \cite{Allers:2013aa} and the references therein. 

\dwnine\ does not look like a source with the young age of TWA or Beta Pic, whose members lack \ion{FeH}{} features around 1\,$\mu$m and have different $H$ and $K$ band shapes. The best match seems to be EROS 0032$-$44, i.e.\ an age of perhaps $\sim$125 Myr (the age of the Pleiades Moving Group is still debated, see \citealt{Famaey:2008aa}), a source which exhibits low-gravity features and lithium absorption \citep{Martin:1999yf}. {This is consistent with the \cite{Allers:2013aa} gravity-sensitive indices that point towards an intermediate gravity (INT-G), indicating an age estimate of roughly 100--{300} Myr. The MagE spectra reported in Burgasser et al. (in prep.) exhibits subtle indications of low surface gravity --- slightly weaker 8183/8195 {\AA} \ion{Na}{i} lines and a modestly stronger 8100 {\AA} VO band than equivalently-classified dwarfs --- but these are not particularly strong features and thus consistent with a source in the few 100 Myr age range.}

\begin{figure}[h!]
\center
\includegraphics[width= \linewidth]{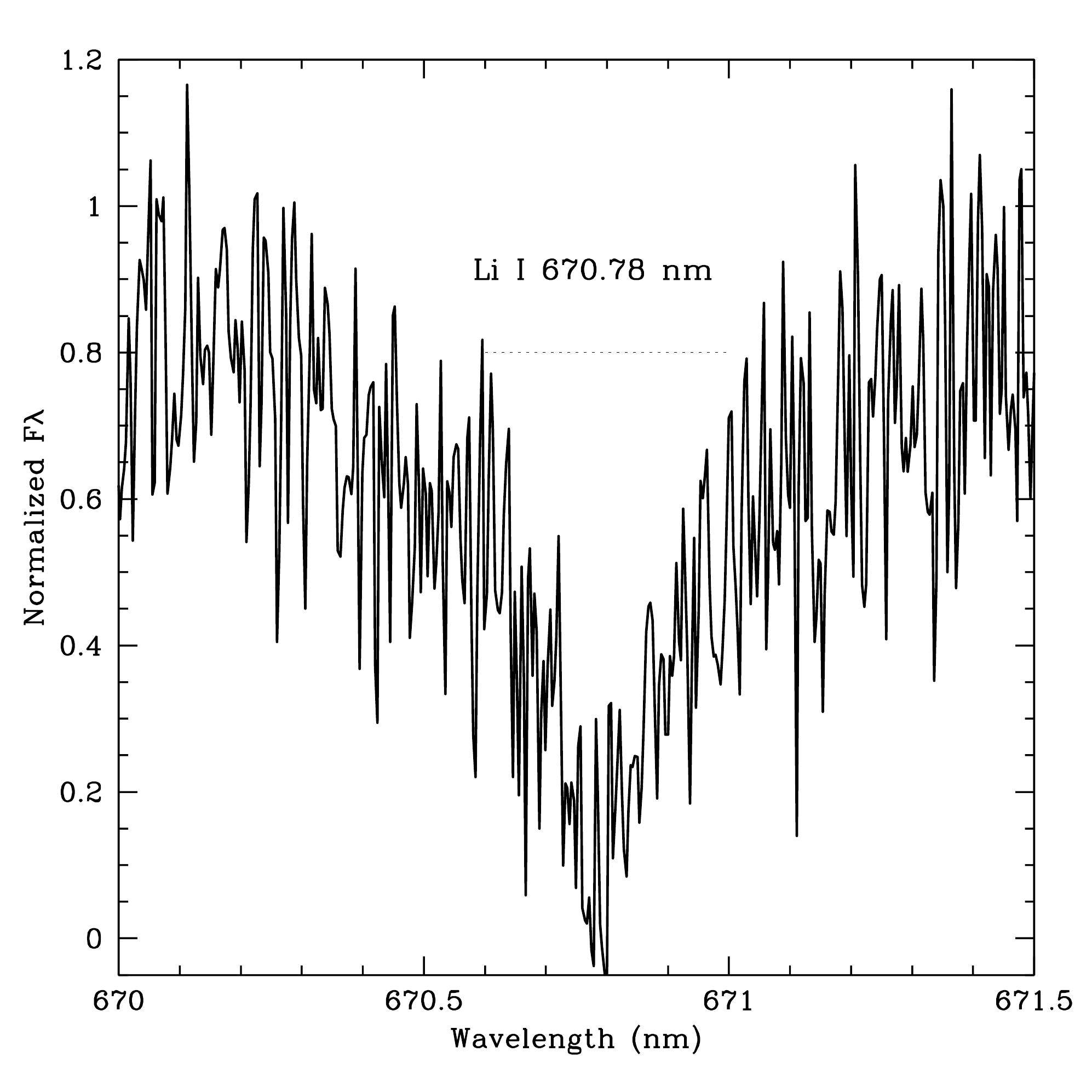}
\caption{The \ion{Li}{i} line in the UVES spectrum of \dwnine. The dotted line marks the integration limits and continuum level chosen to measure the equivalent width.}
\label{fig:UVES_Li}
\end{figure}

The UVES spectrum of \dwnine\ shows a strong \ion{Li}{i} absorption feature. A \ion{Li}{i} equivalent width of $2.4 \pm 0.1\,\AA$ was measured by direct integration of the line profile shown in Fig.~\ref{fig:UVES_Li}. Such a strong \ion{Li}{i} line has already been reported in high-resolution spectra of late-M and L dwarfs, and it indicates that lithium has not been depleted in the primary component of the binary \citep{Pavlenko:2007aa}. 
{The spectral resolution of our UVES spectrum is much higher than the resolution that has been used to classify L dwarfs using gravity sensitive features
\citep{Cruz:2009fk}, and we refrain from putting our data in a classification system
based on low-resolution spectra.  In future work we plan to compare our UVES spectra of the
\cite{Sahlmann:2014aa} targets with high-resolution synthetic spectra to obtain quantitative
estimates of their surface gravities (Mart\'in et al. in prep.).}

As discussed in \citet{Magazzu:1993kx}, the presence of a strong \ion{Li}{i} line coupled with the effective temperature of  \dwnine\ already indicates that it must have a substellar mass lower than about 0.065 solar masses and a relatively young age between 100 and 1000 Myr. A more precise estimate on the age of \dwnine\ using lithium and other age indicators is presented in Sect. \ref{sec:age}. 

\subsection{Spectral binarity}
Unresolved ultracool binaries with late-M/L dwarf and T dwarf components can be discovered and characterised by disentangling the spectral features of individual components in their combined-light spectra (e.g.\ \citealt{Burgasser:2007aa}). For \dwnine, the estimated component masses from \citet{Sahlmann:2013ab} for an age of $\sim$1~Gyr suggested that the secondary of this system could be a T dwarf.  However, we do not see the $H$-band `dip' feature commonly present in combined-light L/T binary spectra, only a slight excess in the 2.05 and 2.2~$\mu$m region. 

We nevertheless applied the methods described in \citet{Burgasser:2010kx} and compared the spectrum of \dwnine\ to binary templates constructed from L and T dwarf spectra in the SpeX Prism Library\footnote{\url{http://www.browndwarfs.org/spexprism}} (SPL, \citealt{Burgasser:2014aa}). This procedure allows us to estimate the spectral types and effective temperatures of the individual components and their $I$-band magnitude difference. 

Because of the \ion{Li}{i} absorption and its near-infrared spectrum, we know that \dwnine\ is relatively young. To reflect this in spectral binary fitting, we {performed the analysis with three sets of templates: those with only `young' sources, only `not young' sources, and with `all' sources}. 
{In each case, this analysis uses binary templates constructed from M9--L5 primaries and L4--T6 secondaries} in the SPL, where individual component spectra were scaled to the absolute $J$-band magnitude -- spectral type relation of \cite{Dupuy:2012fk}. The F test was applied to compare the quality of different fits, which relies on the assumption of Gaussian uncertainties\footnote{Note that the used $\chi^2$ prescription (Eq. (1) in \citealt{Burgasser:2010kx}) does not account for the noise spectrum of the template, because this would bias the result towards the most noisy templates.}. {The young sources} were selected as having previously identified signatures of low surface gravity (e.g.\ \citealt{Cruz:2009fk, Allers:2013aa}), however this sample is dominated by sources older than $\sim$100 Myr. {The `not young' sample excludes these young sources.}

\begin{figure}[h!]
\centering
\includegraphics[width = \linewidth, trim=15mm 0 5mm 0, clip=true]{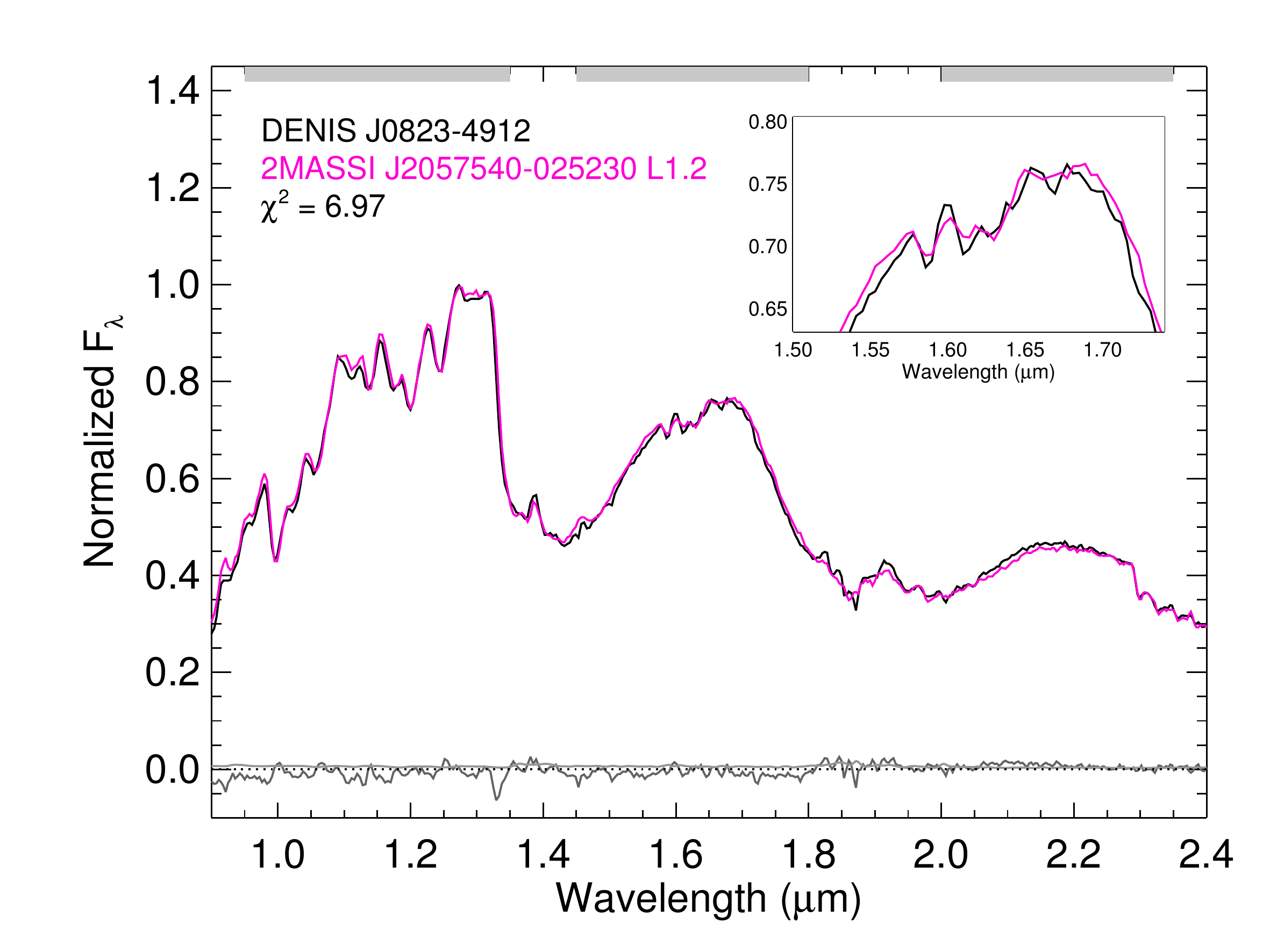}
\caption{{Result of spectrum fitting with single templates. The single template 2M2057$-$02 (blue line) is the best fit to the SpeX spectrum of \dwnine\ (black line) both for the `young' and `all' set of templates.} At the top of the panel, the spectral ranges considered for the fit are indicated by grey horizontal bars, where telluric water absorption dominates the gaps. The inset shows a close-up of the $H$-band region. The residual spectrum is shown in grey.} 
\label{fig:spex2}
\end{figure}

\begin{figure}[h!]
\centering
\includegraphics[width = \linewidth, trim=15mm 0 5mm 0, clip=true]{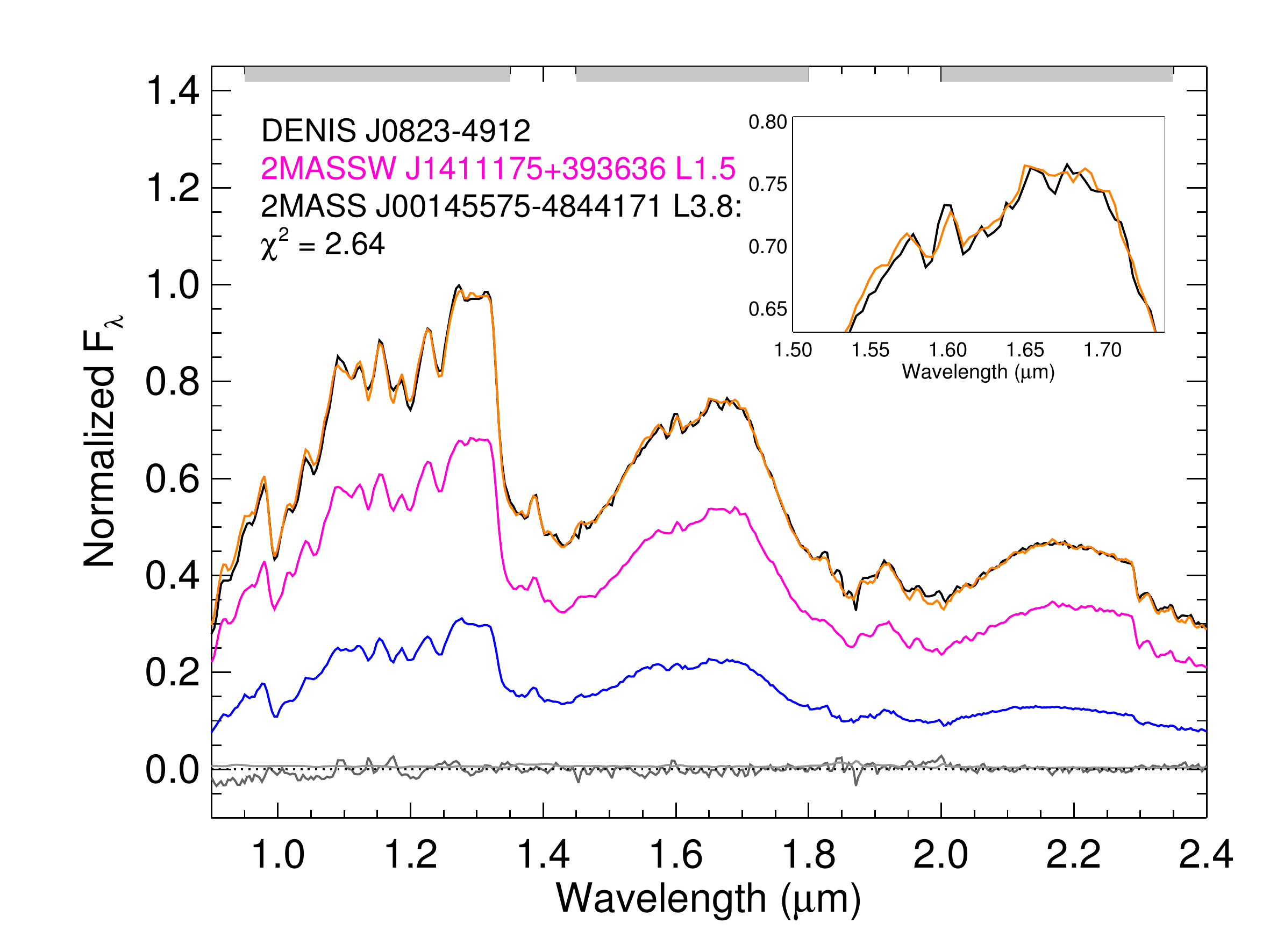}
\includegraphics[width = \linewidth, trim=15mm 0 5mm 0, clip=true]{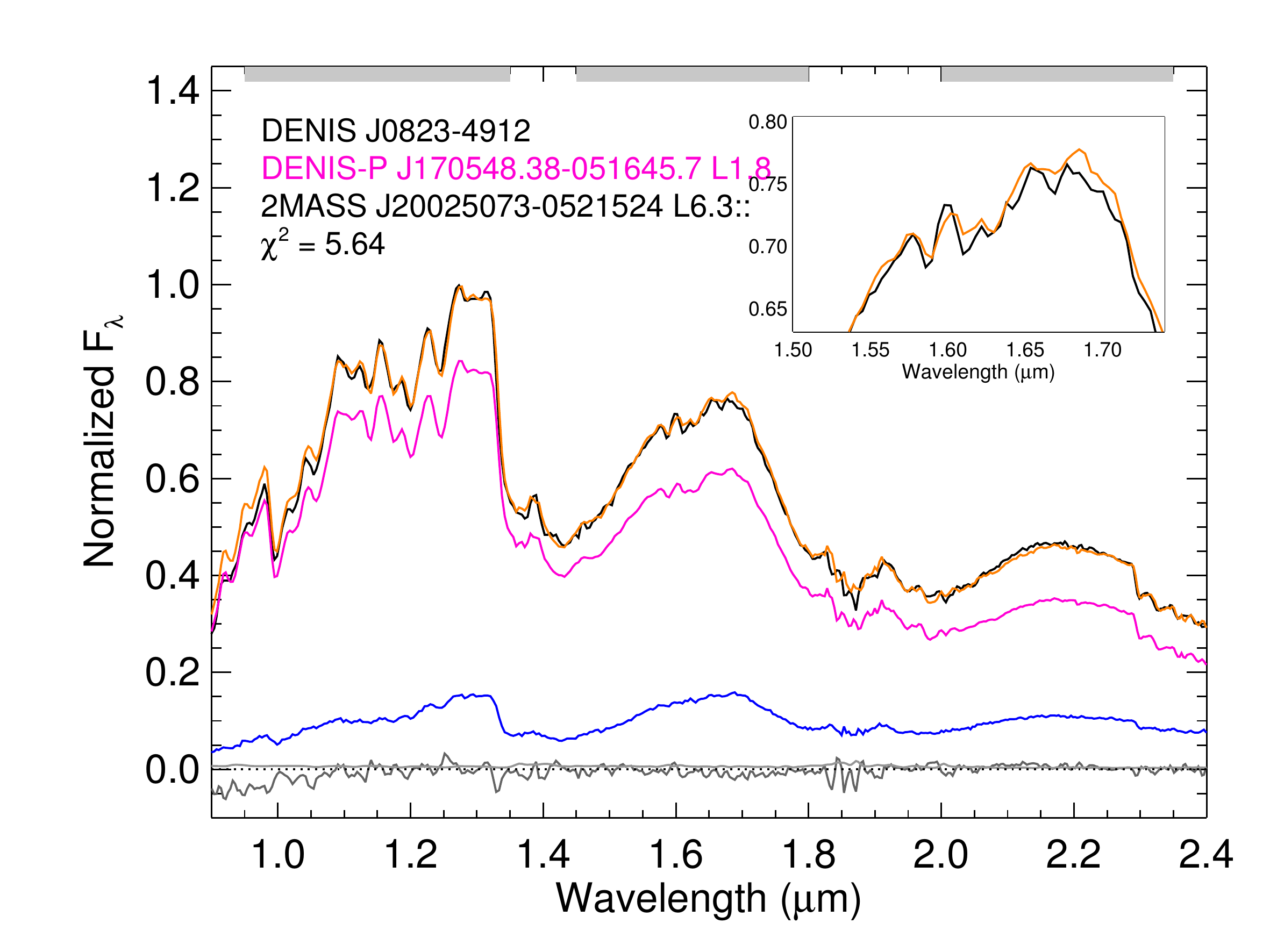}
\caption{{Results of spectrum fitting with binary templates (cf. Fig. \ref{fig:spex2}). \emph{Top}: When using `all' templates, the SpeX spectrum of \dwnine\ (black line) is best fit by the binary template (brown line) that is a combination of 2MASSW J1411175$+$393636 (red line) and 2MASS J00145575$-$4844171 (blue line). \emph{Bottom}: The same when using only `young' templates. The best-fit binary template (brown) is a combination of DE1705$-$05 (red) and 2M2002$-$05 (blue).}}
\label{fig:spex2_normal}
\end{figure}

{For both the `young'  and `all'  sets of templates, the best single match to the spectrum of \dwnine\ is that of the L1.5 \object{2MASS J20575409-0252302} (hereafter 2M2057$-$02; \citealt{Cruz:2003aa,Burgasser:2004aa,Allers:2013aa}), which is shown in Fig. \ref{fig:spex2}.} {The best-fit binary template for the unrestricted template set (`all') is shown in the top panel of Fig. \ref{fig:spex2_normal}. Is is a} combination of the L1.5 \object{2MASSW J1411175+393636} \citep{Kirkpatrick:2000aa} and the L2.5 pec \object{2MASS J00145575-4844171} \citep{Kirkpatrick:2008uk}, where the latter's near-infrared appearance classifies it closer to a spectral type of L4. This provides a significantly better match than the spectrum of 2M2057$-$02, {particularly in the shape of the 2.1 $\mu$m $K$-band peak (likely due to stronger H$_2$ absorption in the secondary), so that} the single template model is rejected with a confidence of 99.9 \% on the basis of the F test. {The influence of the companion on the blended light spectrum is thus clearly detected.}  

{For the `young' template set,} the bottom panel of Fig.~\ref{fig:spex2_normal} shows the best-fit young binary template, a combination of \object{DENIS-P J170548.4-051645.7} (DE1705$-$05, \citealt{Kendall:2004aa}) and \object{2MASS J20025073-0521524} (2M2002$-$05, \citealt{Cruz:2007aa}). The addition of the L6 fills in the excess flux at $K$-band, and this binary template provides a significantly better match than the spectrum of {2M2057$-$02}. The single template model is rejected with a confidence of {75 \%} on the basis of the F test. 

To obtain the parameters of \dwnine\ and their confidence intervals, we combine all template fits and assign them relative weights according to their $\chi^2$ values {to compute weighted average values with uncertainties like in  \citet{Burgasser:2010kx}.} {The results are summarised in Table \ref{table:specfit}, which also lists the number of available binary templates and the F-test confidence. Effective temperatures were derived using the relations of \cite{Looper:2008ab} and include an additional 0.5 spectral subtype uncertainty for optically classified sources.} We measured the 2MASS $J$ magnitudes for both components of the binary templates and used the \cite{Schmidt:2010aa} relation between spectral type and SDSS $i$ -- 2MASS $J$ colour to obtain estimates for the SDSS $i$ magnitude difference between primary and secondary. We found that this is indistinguishable from integrating the Bessel $I$ or the SDSS $i$ bandpasses from the template spectra directly. The uncertainties are dominated by scatter in the viable sources{, thus $\Delta i_\mathrm{SDSS} \simeq \Delta I$. }
{For every individual spectral fit, we also computed bolometric corrections \citep{Liu:2010fk} for the components of the spectral templates and estimated the $K$-band magnitude difference $\Delta K_\mathrm{MKO}$. We used these to compute the component bolometric luminosities $L_\mathrm{bol}$ based on the system's combined light absolute MKO magnitude, obtained from absolute 2MASS $K_\mathrm{s}$ and the MKO--2MASS correction from the spectrum. The uncertainty in $L_\mathrm{bol}$ includes 0.08 {mag} uncertainty in the $K$-band bolometric correction.} 

\begin{table}[h!]
\caption{Results of spectral binary fitting with different sets of templates}    
\label{table:specfit}      
\centering       
\small                
\begin{tabular}{c r c c c} 
\hline\hline    
\multicolumn{2}{c}{} & `young' & `not young' & `all' \\
\hline
\multicolumn{2}{c}{\# Primaries} & 59 & 381& 440\\
\multicolumn{2}{c}{\# Secondaries} & 28 & 341& 369\\
\multicolumn{2}{c}{\# Binaries} & 1\,634 & 128\,868 &161\,121\\
\multicolumn{2}{c}{Primary SpT} & L$1.5\pm0.6$& L$1.1\pm0.6$& L$1.1\pm0.6$\\
\multicolumn{2}{c}{Secondary SpT} & L$5.5\pm1.1$& L$4.2\pm1.4$& L$5.3\pm2.3$\\
\multicolumn{2}{c}{Primary $T_\mathrm{eff}$ (K)} & $2150\pm 100$& $2190\pm 90$& $2200\pm 90$\\
\multicolumn{2}{c}{Secondary $T_\mathrm{eff}$  (K)} & $1670\pm 140$& $1810\pm 180$& $1720\pm 230$\\
\multicolumn{2}{c}{Primary $\log\!L_\mathrm{bol}$  ($L_\sun$)} & $-3.74\pm 0.04$& $-3.77\pm 0.07$& $-3.73\pm 0.08$\\
\multicolumn{2}{c}{Second. $\log\!L_\mathrm{bol}$  ($L_\sun$)} & $-4.20\pm 0.08$& $-4.15\pm 0.14$& $-4.25\pm 0.20$\\
\multicolumn{2}{c}{$ \Delta I $  (mag)} & $2.4\pm0.6$& $1.7\pm0.9$& $2.3\pm1.2$\\
\multicolumn{2}{c}{F-test conf. (\%)} & 75 & 99.9 & 99.9\\
\hline
\end{tabular}
\end{table}

In all cases, the spectral types of the primary and secondary component are consistent with L1.5 and L5, respectively. All other properties are also derived consistently within their uncertainties for different template sets. For the `not young' and `all' sets, the single template model is rejected with a confidence of 99.9 \%, which we interpret as a significant detection of the companion in the blended light spectrum. For the `young' templates, the rejection confidence is 75 \%, which would be considered insufficient evidence of (young) multiplicity based on spectroscopy alone.  However, this statistic is skewed by the small number of young templates available in the SPL, and spectral binary `significance' is less relevant since we already know that this system is a binary. We can still use the F-statistic to weight the spectral types from this analysis, which are in agreement with the `not young' and `all' templates samples. {The high significance of the F test with `all' templates may suggest that spectral binary fitting can be a viable procedure for detecting L+L spectral type binary systems. We caution, however, that the binarity of \dwnine\ was established with astrometry, which simplifies the  interpretation of the spectral fitting results.} 

Because we know that \dwnine\ is relatively young, we adopt the parameters determined with the young templates. {This yields component spectral types of {L1.5$\pm$0.6 and L5.5$\pm$1.1} and effective temperatures of $2150\pm 100$ K and $1670\pm 140$ K for the primary and secondary components of \dwnine, respectively. The corresponding} magnitude difference of $ \Delta I = 2.4\pm0.6$ is valid for the passband of our FORS2 observations.   

\subsection{Constraints on age and physical parameters}\label{sec:age}
\begin{figure}[h!]
\centering
\includegraphics[width = \linewidth]{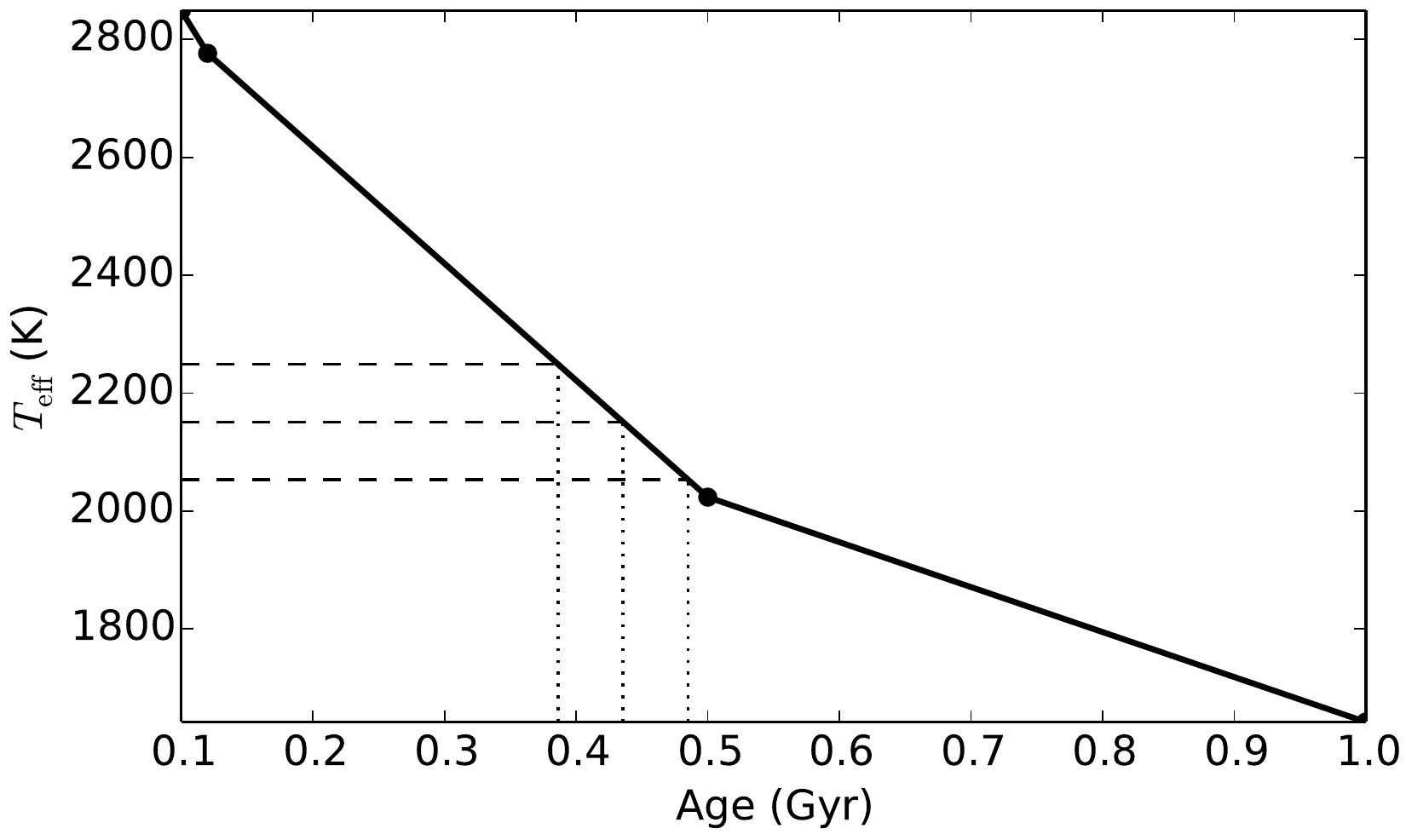}
\caption{Upper age limit from the presence of \ion{Li}{i} absorption in the primary's spectrum. {The curve shows the effective temperature of objects that have depleted 90 \% of their lithium (i.e.\ lithium abundance = 0.1) as a function of age after interpolation of the DUSTY models. Horizontal dashed lines indicate the primary's effective temperature range, which translates into an upper age limit of $\sim$0.5 Gyr.}}
\label{fig:lithiumage}
\end{figure}

The determination of individual effective temperatures in the previous section allows us to employ models of (sub-)stellar evolution to set constraints on the system's age and its properties. The first comes from the detection of \ion{Li}{i} absorption in the spectrum, which implies a mass of $\lesssim0.065 \,M_\sun$ for the primary. Figure \ref{fig:lithiumage} shows how this translates into an upper age limit of {$\sim$0.5 Gyr} when coupled to the primary's effective temperature and the DUSTY \citep{Chabrier:2000kx} evolutionary models.

\begin{figure}[h!]
\centering
\includegraphics[width = \linewidth]{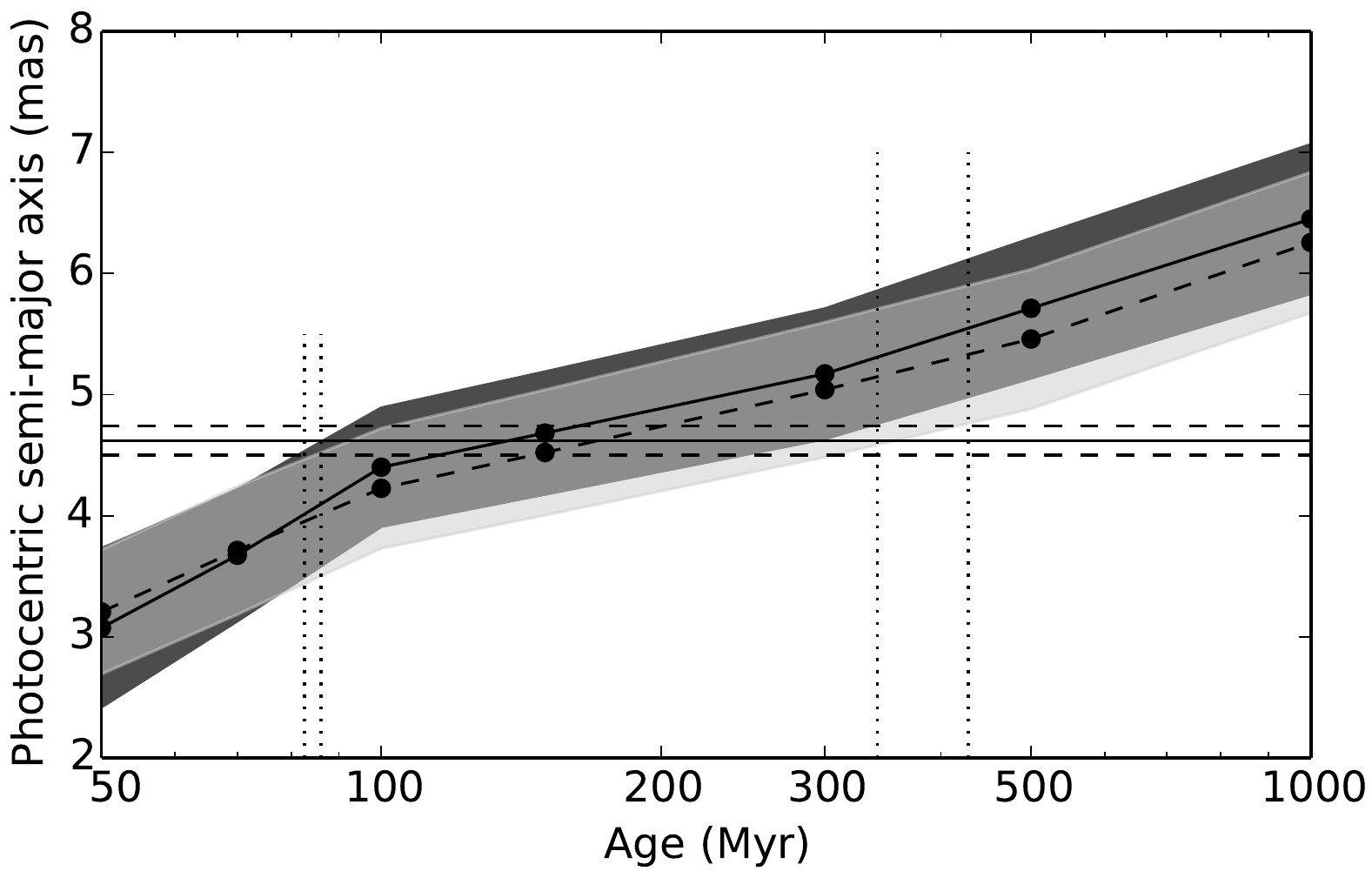}
\caption{Age constraints derived on the basis of two sets of evolutionary models, {derived using effective temperature -- mass relationships}. The curves shows the photocentric orbit size $\alpha_{C}$ for various ages and fixed effective temperatures, where the solid and dashed {curve corresponds to the DUSTY and \cite{Saumon:2008vn} models, respectively. The shaded regions delineate the uncertainties (dark grey for DUSTY, light grey for \cite{Saumon:2008vn}). The horizontal lines show the measured value of $\alpha$ and its uncertainty. Vertical dotted lines indicate age limits derived from  the photocentric orbit.}}
\label{fig:age2}
\end{figure}

\begin{figure}[h!]
\centering
\includegraphics[width = \linewidth]{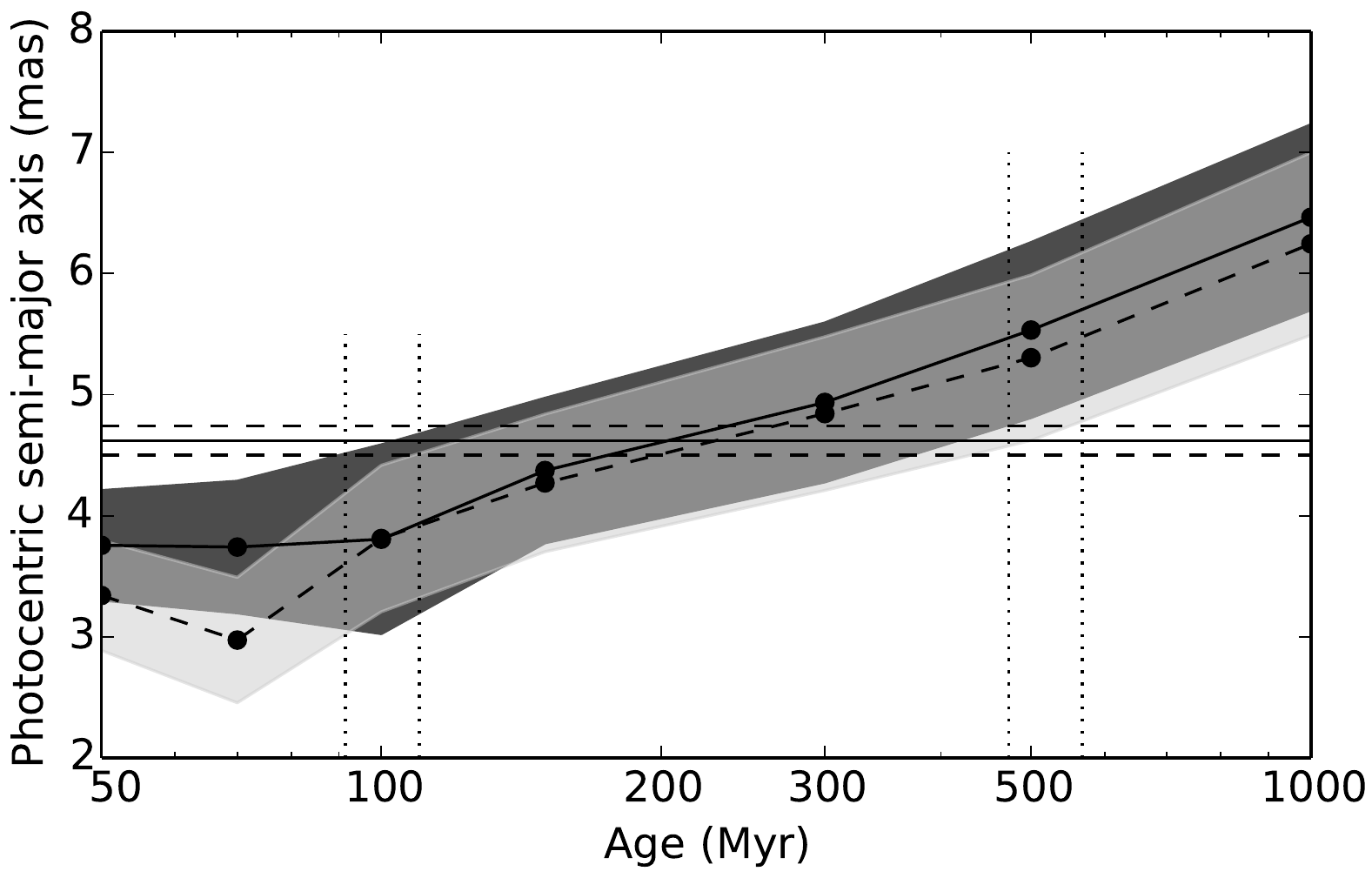}
\caption{{Same as Fig. \ref{fig:age2}, but derived using the bolometric luminosity -- mass relationships.}}
\label{fig:age3}
\end{figure}

Additional constraints can be derived from the photocentric orbit, which was measured in the $I$-band (I\_Bessel filter of FORS2, centred at 760 nm). Because of the moderate magnitude difference between the two components of \dwnine, we have to take into account the light contribution of the companion if we want to determine the barycentric orbit size that relates to the system's mass function. 
The fractional mass
\begin{equation}\label{eq:1}
f = M_2 / (M_1+M_2)
\end{equation}
and the fractional luminosity in the observation passband
\begin{equation}\label{eq:2}
\beta = L_2 / (L_1+L_2) = (1+10^{0.4\ \Delta m})^{-1},
\end{equation}
where $\Delta m$ is the magnitude difference, define the relationship between the semimajor axis $\alpha$ of the photocentre orbit and the semimajor axis $a_{rel}$ of the relative orbit, both measured in mas:
\begin{equation}\label{eq:3}
\alpha = a_{rel} \, (f-\beta).
\end{equation}
An independent constraint on the relative semimajor axis is given by Kepler's law:
\begin{equation}\label{eq:4}
G\, (M_1+M_2) = 4\, {\rm \pi}^2 \frac{\bar a_{rel}^3}{P^2},
\end{equation}
were $G$ is the gravitational constant, $\bar a_{rel}$ is measured in metres and $P$ is in seconds. The relation between $\bar a_{rel}$ and $a_{rel}$ is given by the parallax. Any combination of the unknown parameters $M_1$, $M_2$, and $\Delta m_I$ has to satisfy Eqs. \ref{eq:3} and \ref{eq:4}, which are constrained by the measured quantities $P$, $\alpha$, and the parallax $\varpi$. {We used the DUSTY and \cite{Saumon:2008vn} models} to impose theoretical relationships between effective temperature {or bolometric luminosity} and mass as a function of age, which is the parameter we want to constrain.

For every accepted binary template in the previous section, we derived theoretical component masses on the basis of the effective temperatures and combined them with the estimated $I$-band magnitude difference and Eqs. \ref{eq:1}--\ref{eq:4} to compute the expected photocentric orbit sizes for a range of ages. For every age, we determined the effective orbit size $\alpha_{C}$ and its uncertainty from the weighted average and the weighted standard deviation, respectively, where again we used the weights corresponding to the $\chi^2$ values of the spectral binary fits. When needed, the models were interpolated linearly in effective temperature, mass, or age. 

\begin{table*}[ht!]
\caption{{Derived age, masses, and mass ratio of \dwnine\ as a function of model and the relationship used to obtain theoretical masses, indicated in the column R. There are two rows for every configuration, which give the parameter ranges as delineated by vertical lines in Figs. \ref{fig:age2} and \ref{fig:age3} and the best-fit values with uncertainties that reflect these ranges.}}    
\label{table:masses}      
\centering                       
\begin{tabular}{c c c c c c c} 
\hline\hline    
Model & R& Age & $M_1$& $M_2$ & $q=M_2/M_1$ \\ 
            &              &(Myr) & ($M_\sun$)& ($M_\sun$)& \\
\hline
DUSTY & $L_\mathrm{bol}$ & [91,473] & [0.028,0.060] & [0.018,0.044] & [0.65,0.74] \\[1pt]
DUSTY & $L_\mathrm{bol}$ & $216^{+257}_{-124}$ & $0.041^{+0.019}_{-0.013}$ & $0.029^{+0.015}_{-0.011}$ & $0.71^{+0.03}_{-0.06}$ \\[1pt]
SM08\tablefootmark{a} & $L_\mathrm{bol}$ & [110,568] & [0.031,0.063] & [0.020,0.045] & [0.65,0.71] \\[1pt]
SM08 & $L_\mathrm{bol}$ & $241^{+326}_{-132}$ & $0.044^{+0.019}_{-0.013}$ & $0.030^{+0.015}_{-0.010}$ & $0.68^{+0.03}_{-0.03}$ \\[3pt]
SM08 & $L_\mathrm{bol}$\tablefootmark{b} & $364^{+497}_{-192}$ & $0.042^{+0.018}_{-0.013}$ & $0.029^{+0.014}_{-0.009}$ & $0.69^{+0.03}_{-0.03}$ \\[3pt]

DUSTY & $T_\mathrm{eff}$ & [83,342] & [0.034,0.054] & [0.022,0.039] & [0.64,0.73] \\[1pt]
DUSTY & $T_\mathrm{eff}$ & $139^{+203}_{-56}$ & $0.041^{+0.013}_{-0.006}$ & $0.029^{+0.011}_{-0.007}$ & $0.70^{+0.03}_{-0.06}$ \\[1pt]
SM08 & $T_\mathrm{eff}$ & [86,428] & [0.035,0.060] & [0.022,0.043] & [0.64,0.71] \\[1pt]
SM08 & $T_\mathrm{eff}$ & $178^{+250}_{-92}$ & $0.045^{+0.015}_{-0.010}$ & $0.030^{+0.012}_{-0.008}$ & $0.68^{+0.03}_{-0.04}$ \\[2pt]
\hline
\end{tabular}
\tablefoot{
\tablefoottext{a}{\cite{Saumon:2008vn}.}
\tablefoottext{b}{These results were obtained when considering a systematic offset in the model that corresponds to making a brown dwarf of a given mass 0.3 dex brighter in terms of $\log\!L_\mathrm{bol}$.}
}
\end{table*}

Figure \ref{fig:age2} shows the estimated orbit size $\alpha_{C}$ a function of age. Using the DUSTY models results in slightly larger values compared to the \cite{Saumon:2008vn} models. The measured orbit size $\alpha$ yields an additional constraint on the system age. {To address the effects of potential systematic differences between spectral type -- effective temperate relations for young and normal sources, which could be relevant in the case of \dwnine, we repeated this analysis using the bolometric luminosities to obtain theoretical masses, leading to similar results, see Figure \ref{fig:age3}.}

{Table \ref{table:masses} lists the numerical results for all configurations, which give an overall consistent picture of \dwnine\ with an age of $\sim$100--500 Myr, primary mass of $M_1=0.028-0.063\,M_\sun$, companion mass of $M_2=0.018-0.045\,M_\sun$, and a mass ratio $q=M_2/M_1= 0.66-0.74$. On average, the bolometric luminosity method yields a slightly older age range, but all results agree well with the upper age limit of $\sim$500 Gyr set by the \ion{Li}{i} detection. 
A typical example of best-fit values is an age of $240^{+330}_{-130}$ Myr, masses of $M_1=0.044^{+0.019}_{-0.013}$ and $M_2=0.030^{+0.015}_{-0.010}$, and a mass ratio of  $q=0.68^{+0.03}_{-0.03}$ (obtained using \cite{Saumon:2008vn} and bolometric luminosities.)} {These age estimates are consistent with the INT-G surface gravity classification and the quality of gravity-sensitive features present in the near-infrared and optical spectra.}

{Finally, the age and mass parameters that we derived for \dwnine\ fall into a range where the results of \cite{Dupuy:2014aa} suggest that theoretical models underpredict the luminosities of brown dwarfs. To quantify the effect that such systematic errors would have on our results, we introduced an 0.3 dex offset in the \cite{Saumon:2008vn} bolometric luminosities that would make a BD with given mass and age $\approx$2 times brighter in $L_\mathrm{bol}$, an amplitude that corresponds to observations \citep{Dupuy:2009fr}. When including the offset, the resulting acceptable age range of \dwnine\ is shifted to 150--860 Myr, but the component masses remain the same, see Table \ref{table:masses}. Because an upper age limit is set by the \ion{Li}{i} detection, we conclude that such a systematic model error has negligible effect on the age and mass properties we derive for \dwnine.}

\subsection{Rotation} 
\begin{figure}[h!]
\center
\includegraphics[width= \linewidth]{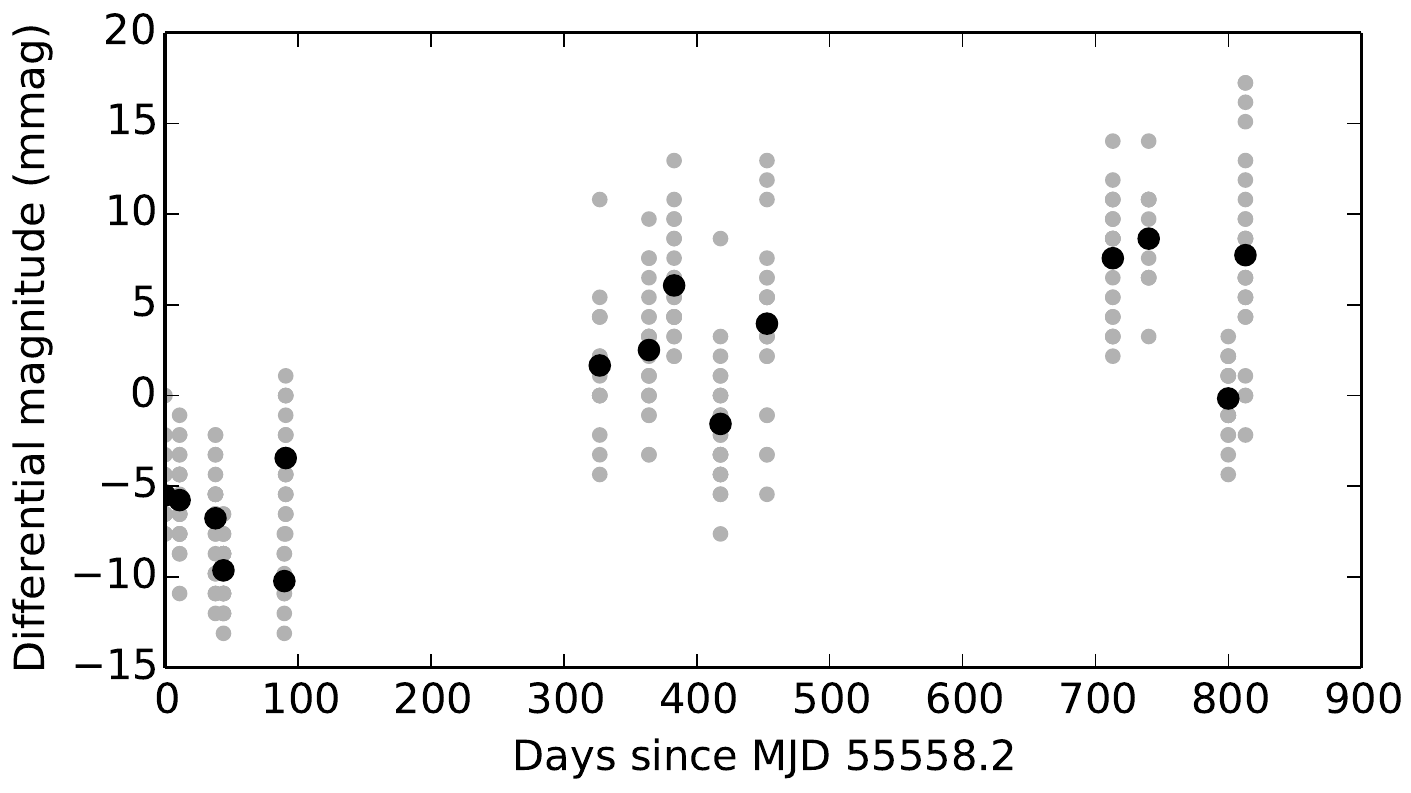}
\caption{Differential magnitude variation of \dwnine\ in $I$-band as a function of time. Grey symbols correspond to measurements in individual FORS2 frames, whereas black circles show the epoch average.}
\label{fig:variability}
\end{figure}

Using the NIRSPEC infrared spectra, we measured a mean projected rotational velocity of $v \sin i = 32\pm4$ km\,s$^{-1}$, which is comparable to that found for other field L1.5 dwarfs \citep{Reiners:2008cr}. If the binary orbit orientation ($\sin i$\,=\,0.79) coincides with the spin axis of the primary, the actual rotational velocity is $\sim$$ 40$~km\,s$^{-1}$. We performed a periodogram analysis of the optical variability of \dwnine\ shown in Fig. \ref{fig:variability}, but found no evidence for periodic variations at correspondingly short rotation periods ($\approx$3 h for a Jupiter-sized body). We only noticed a slight variation of $\sim$10 mmag over 800 days.

\subsection{Radial velocity orbit}
\begin{figure}[h!]
\center
\includegraphics[width= \linewidth]{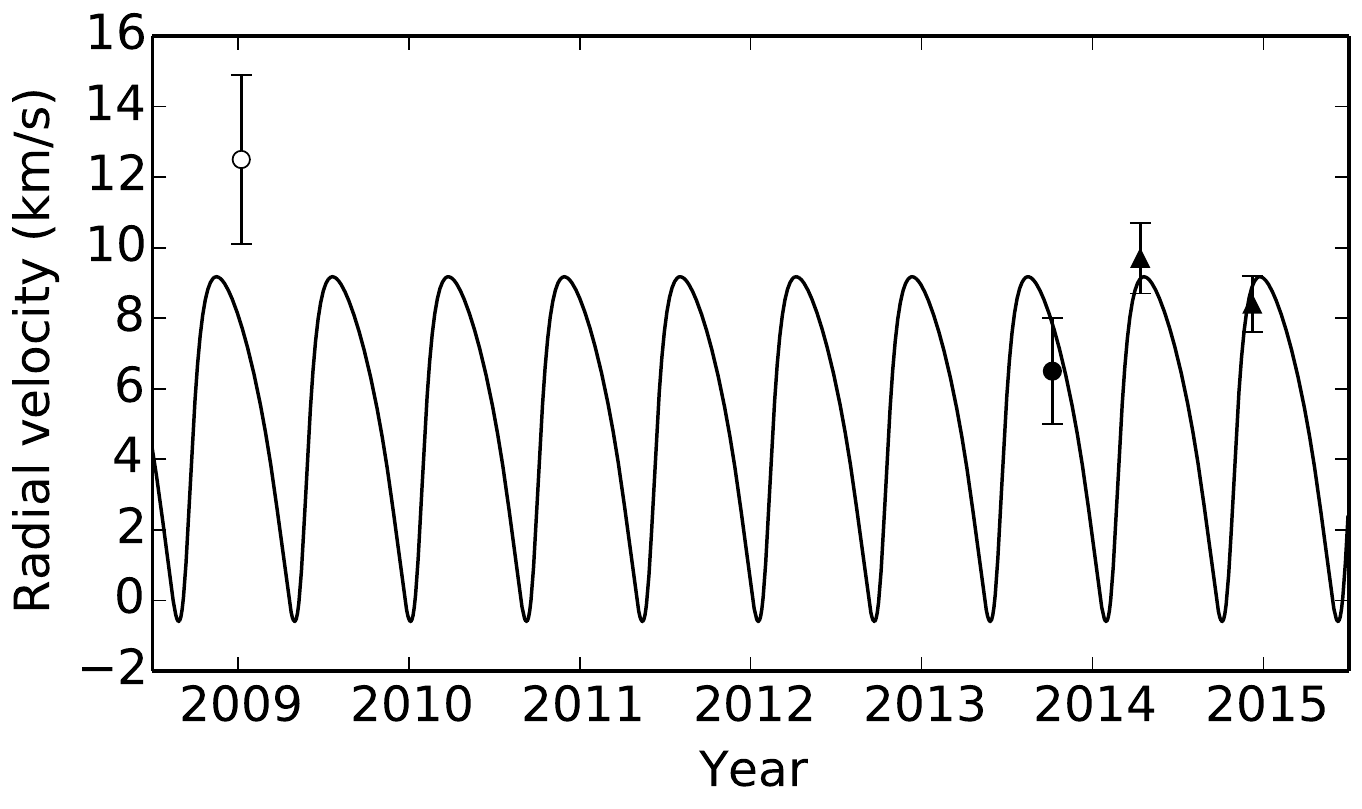}
\includegraphics[width= \linewidth]{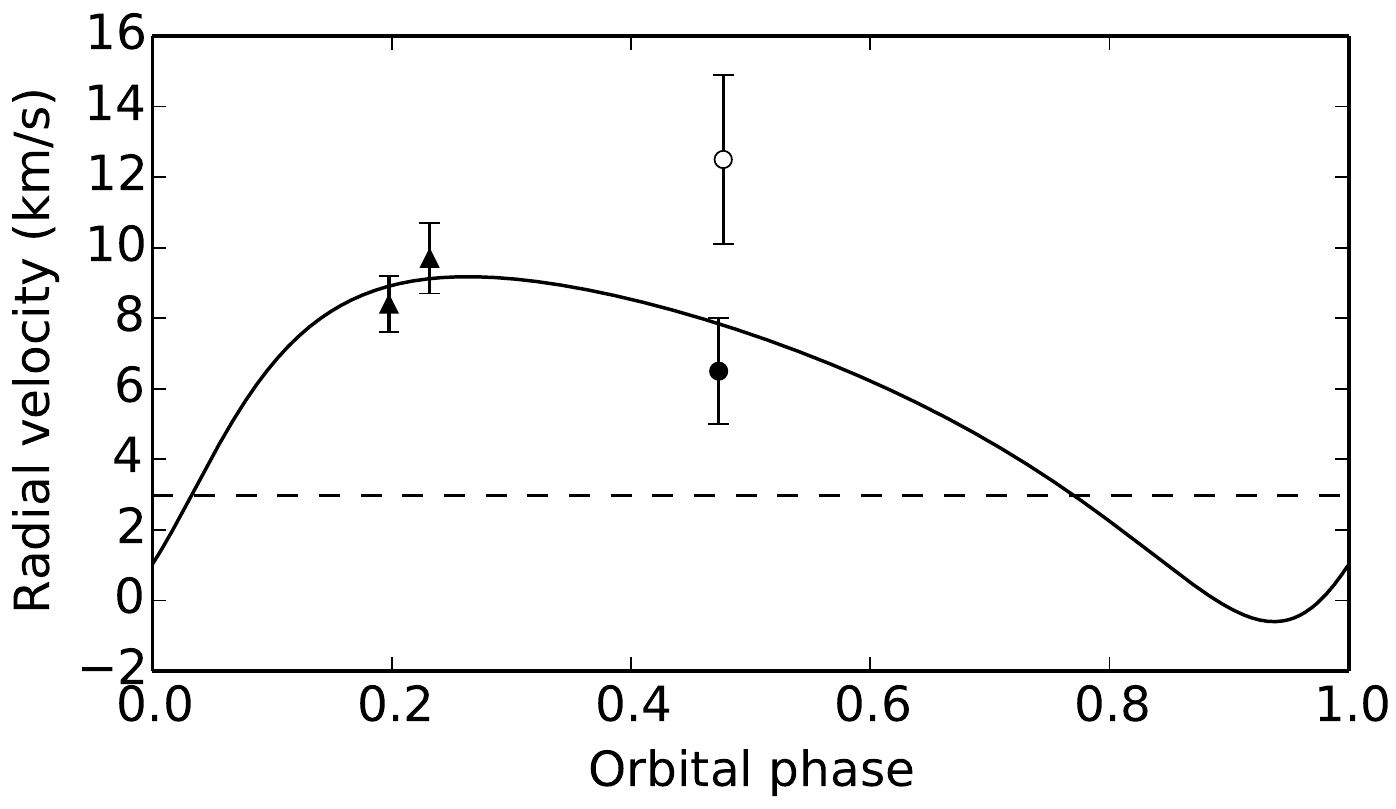}
\caption{Estimated radial velocity curve of \dwnine A ($M_1=0.044\, M_\sun$, $M_2=0.030\, M_\sun$) as a function of time (top panel) and orbital phase (bottom panel, phase 0 = periastron). The measurements with UVES (solid circle), NIRSPEC (triangles), and MagE (open circle) are shown and the systemic velocity is indicated by the dashed line. } 
\label{fig:RV}
\end{figure}

Figure \ref{fig:RV} shows the radial velocity measurements (see Table \ref{table:3}) and the expected radial velocity curve of \dwnine A shifted by the systemic velocity  $\gamma \simeq 2.98 \pm 0.75$ km\,s$^{-1}$ that we modelled as a constant offset to the measurements. We assumed masses of $M_1=0.044\, M_\sun$ and $M_2=0.030\, M_\sun$, but the curve is essentially the same for a younger {or older} configuration, e.g.\ $M_1=0.031\, M_\sun$ and $M_2=0.020\, M_\sun$. The available measurements are compatible with the shown curve, which allows us to conclude that we resolved the ambiguity of 180\degr\ in inclination and longitude of ascending node, which means that the values reported in Table \ref{table:1} are accurate. More radial velocity measurements {would be} needed to independently constrain the system parameters.

\begin{table}
\caption{Radial velocity measurements of \dwnine A}    
\label{table:3}      
\centering                       
\begin{tabular}{r r r r r} 
\hline\hline    
Instr. &Date& Epoch & RV & $v \sin{i}$\\
 & & (MJD)& (km/s)& (km/s)\\
\hline
MagE & 2008-01-08& 54838.21756 &  $12.5 \pm 2.4$  & $\cdots$ \\
{\small UVES} &2013-10-07 & 56572.33321&  $6.5 \pm 1.5$  & $26 \pm 5$\\
{\small NIRSPEC}&2014-04-13&     56760.22524  & $9.7\pm1.0$ &   $34\pm5$\\
{\small NIRSPEC}&2014-12-08  &   56999.55533 &  $8.4\pm0.8$  &  $29\pm6$\\

\hline
\end{tabular}
\end{table}

\section{Discussion}
Young sources in the solar neighbourhood are often found in associations that have similar kinematic properties. Since all relevant quantities, i.e.\ coordinates, parallax, proper motions, and radial velocity, are now known for \dwnine, we examined its potential membership in a young association using the \textsc{\small BANYAN II} web tool \citep{Gagne:2014aa}. According to \textsc{\small BANYAN II}, \dwnine\ has a 100 \% probability of belonging to the field population. The probability of belonging to one of the 7 young associations probed is reported as 0.00 \%.

\dwnine\ thus appears isolated in terms of membership, although there may exist numerous very-low-mass few-member young associations that are not recognised yet in the solar neighbourhood. Searches for young very-low-mass stars outside star-forming regions indicate that they are typically near larger star-forming regions but not inside them \citep{Martin:1996aa, Valdivielso:2009aa}.

Another consequence of the new age estimate is related to the mass ratio  ($q=M_2/M_1$) of the \dwnine\ binary that had a reported value of 0.36 derived under the assumption of a system age of $\sim$1 Gyr \citep{Sahlmann:2013ab}. With the characterisation presented here, the primary mass is lower but the mass ratio is higher in the range of $\sim${0.66--0.74}. This system appears now more similar to the bulk of known ultracool dwarf binaries, whose mass ratio distribution is dominated by systems with $q>0.7$ \citep{Burgasser:2007ix}. {However, its mass ratio is still uncommon and lies in the tail of the $q$-distribution of field brown dwarf binaries \citep[e.g.][]{Liu:2010fk}}. The estimated companion mass in this binary system remains in the range of $\sim$30 Jupiter masses, which corresponds to the most massive planets found around Sun-like stars \citep{Sahlmann:2011fk}. 

\section{Conclusions}
We have presented the detailed characterisation of the tight brown dwarf binary system \dwnine. After the orbit discovery with optical ground-based astrometry, we obtained follow-up optical and infrared spectroscopy. The combination of spectral modelling with age indicators and evolutionary models leads us to the following conclusions:
\begin{itemize}
  \item \dwnine\ is a system composed of two brown dwarfs with spectral types of {L1.5$\pm$0.6 and L5.5$\pm$1.1} that orbit each other in $\sim248$ days. 
  \item {The spectrum of this system is significantly better matched to a binary template than a single spectral template, providing an indirect detection of the companion in the near-infrared. Based on the classification of the templates, we infer the components' effective temperatures to be {$2150\pm100$ K and $1670\pm140$ K}.}
  
  \item We estimate the age of the system to be in the range of{ 80--500 Myr}, which is supported by the detection of \ion{Li}{i} absorption in the optical spectrum and by the properties of the near-infrared spectrum. Refined upper and lower age bounds were derived from the measured photocentric orbit size and theoretical relationships between mass and effective temperature {or bolometric luminosity}.
  \item \dwnine\ does not appear to belong to any of the known nearby young associations.
  \item Evolutionary models predict component masses in the ranges of {$M_1\simeq0.028-0.063\,M_\sun$ and $M_2\simeq0.018-0.045\,M_\sun$ with a mass ratio of $q=0.64-0.74$}. Both components are thus substellar objects below the lithium-burning mass limit and may have masses comparable to the most massive planets found around Sun-like stars. 
\end{itemize}
At a distance of 20.7 pc, \dwnine\ is a rare example of a nearby brown dwarf binary with well-characterised orbit, component properties, and age.

\begin{acknowledgements}
J.S. is supported by an ESA Research Fellowship in Space Science. A.J.B. acknowledges support by the Tri-continental Talent programme (CEI Canarias: Campus Atl\'antico Tricontinental). We thank our referee, T. Dupuy, for a thorough review of our work. This research made use of the databases at the Centre de Donn\'ees astronomiques de Strasbourg (\url{http://cds.u-strasbg.fr}), NASA's Astrophysics Data System Service (\url{http://adsabs.harvard.edu/abstract\_service.html}), the paper repositories at arXiv, the M, L, T, and Y dwarf compendium housed at \url{DwarfArchives.org}, the SpeX Prism Spectral Libraries at \url{http://www.browndwarfs.org/spexprism}, and of Astropy, a community-developed core Python package for Astronomy \citep{Astropy-Collaboration:2013aa}. Portions of the data presented herein were obtained at the W. M. Keck Observatory, which is operated as a scientific partnership among the California Institute of Technology, the University of California and the National Aeronautics and Space Administration. The Observatory was made possible by the generous financial support of the W. M.~Keck Foundation. The authors wish to recognise and acknowledge the very significant cultural role and reverence that the summit of Mauna Kea has always had within the indigenous Hawaiian community.  We are most fortunate to have the opportunity to conduct observations from this mountain.
\end{acknowledgements}

\bibliographystyle{aa} 
\bibliography{/Users/sahlmann/astro/papers} 

\end{document}